\documentclass[%
preprint,
superscriptaddress,
 amsmath,amssymb,
 aps,
]{revtex4-2}

\usepackage{graphicx}
\usepackage{dcolumn}
\usepackage{bm}
\usepackage{lipsum}
\usepackage{mathtools}
\usepackage{amsmath}
\usepackage{enumitem}
\usepackage{caption}
\usepackage{subcaption}
\usepackage{tikz-cd}
\usepackage[normalem]{ulem}
\usepackage{hyperref}
\usepackage[capitalise]{cleveref}

\preprint{APS/123-QED}
\begin{document}

\title{Inversion Domain Boundaries in Wurzite GaN}

\author{M. M. F. Umar}
    \email{mfu102@psu.edu}
    \affiliation{Department of Physics, The Pennsylvania State University}
\author{Jorge O. Sofo}%
    \email{sofo@psu.edu}
    \affiliation{Department of Physics,
	The Pennsylvania State University}
    \affiliation{Department of Materials Science and Engineering, and Materials Research Institute\\
The Pennsylvania State University}
    
\date{\today}

\begin{abstract}
We present two models for the atomic structure of inversion domain boundaries in wurzite GaN, that have not been discussed in existing literature. 
Using density functional theory, we find that one of these models has a  lower formation energy than a previously proposed model known as Holt-$IDB$. 
Although this newly proposed model has a formation energy higher that the accepted lower energy structure, known as $IDB^*$, we argue that it can be formed under typical growth conditions.
We present evidence that it may have been already observed in experiments, albeit misidentified as Holt-$IDB$. 
Our analysis was facilitated by a convenient notation, that we introduced, to characterize these models; it is based on the mismatch in crystal stacking sequence across the $\{10\overline{1}0\}$ plane. 
Additionally, we introduce an improved method to calculate energies of certain domain walls that challenge the periodic boundary conditions needed for plane-wave density functional theory methods.
This new method provides improved estimations of domain wall energies.
\begin{description}
\item[Keywords]
GaN, inversion domain boundary, wurzite crystal structure. 
\end{description}
\end{abstract}

\maketitle
\section{Introduction\label{sec:Introduction}}
Wide-bandgap semiconductors such as Gallium Nitride (GaN) and other III-Nitrides have garnered significant research interest due to their potential in a wide range of applications such as optoelectronics~\cite{service_nitrides_2010,li_gan_2012} and high temperature/power electronics~\cite{trew_high_1997,mishra_gan_1998,ikeda_gan_2010,kuzuhara_algan/gan_2016}. 
GaN based nano-structures for device applications are typically fabricated using metal-organic vapor deposition (MOVCD)~\cite{deb_faceted_2005,hersee_controlled_2006,chen_homoepitaxial_2010,bergbauer_continuous-flux_2010} or molecular beam epitaxy (MBE)~\cite{sekiguchi_ti-mask_2008,bengoechea-encabo_understanding_2011, geelhaar_properties_2011}. 
The estimation of formation energies of defects appearing during epitaxial growth provides guidance into the desired process of growth and helps rationalize the resulting morphology. 
At the same time, the changes induced by the defects on the electronic states of the system have an impact on device performance. 
The electronic signature of defects are determined using  ab initio calculations on atomic models of said defects. 
Due to their extended nature planar defects are an important category of defects in GaN. 
In the $\{10\overline{1}0\}$ plane, defects that extend throughout GaN epitaxial films have been observed in transmission electron microscopy (TEM) experiments~\cite{sitar_structural_1989,yeadon_surface_1997,smith_characterization_1995}.
These planar defects can be explained by stacking mismatch boundaries or inversion domain boundaries. 
Due to the fact that stacking mismatch boundaries can be easily terminated by basal stacking faults and also inversion domain boundaries (IDB) can have much lower formation energies, Northrup et al.~\cite{northrup_inversion_1996} have shown that a particular IDB model, called $IDB^*$, is the most suitable configuration for planar defects in the $\{10\overline{1}0\}$ plane.\\

An IDB is defined as a $\{10\overline{1}0\}$-prismatic plane separating two domains of the GaN crystal, where the spontaneous polarization of the abutting domains have opposite directions. 
In Northrup's $IDB^*$ model~\cite{northrup_inversion_1996}, the Gallium (Ga) and Nitrogen (N) sub-lattice locations are interchanged on one side of the  $\{10\overline{1}0\}$ defect plane \emph{and} the domains have a $\tfrac{1}{2}\mathbf{c}$ relative shift between them. 
The interchange of Ga and N sub-lattices flips the direction of the spontaneous polarization along $[0001]$.
The $\tfrac{1}{2}\mathbf{c}$ relative shift is introduced to avoid the energetically unfavorable Ga-Ga and N-N wrong bonds, where $\mathbf{c}$ is the conventionally denoted primitive lattice vector of a hexagonal close packed lattice. 
The $IDB^*$ model is electronically inert~\cite{northrup_inversion_1996}, which  aligns well with the experimental observations that GaN samples exhibit consistent luminescence characteristics despite the presence of a high density of extended defects $(10^{10}\,cm^{-2})$~\cite{lester_high_1995}. 
The existence of the $IDB^*$ configuration has been verified experimentally~\cite{cherns_determination_1998,potin_1010_1999,potin_evidence_1999}. 
The defect configuration resulting from the species sub-lattice interchange on one side of a $\{10\overline{1}0\}$ plane, without the $\tfrac{1}{2}\mathbf{c}$ relative shift is referred to as the Holt-$IDB$ model~\cite{holt_antiphase_1969}. 
Due to the presence of the wrong bonds the Holt-$IDB$ model has a higher domain wall energy and is electronically active, i.e. it produces states in the band-gap. 
Nevertheless, experimental observations of Holt-$IDB$ are also found in the literature~\cite{potin_1010_1999,potin_evidence_1999}.\\

In this paper, we propose two new models for IDBs. 
We refer to these models as $IDB^{\prime}$ and $IDB^{\prime\prime}$. 
Of these, the $IDB^{\prime}$ has a domain wall energy higher than $IDB^*$ but lower than Holt-$IDB$. 
We will show that $IDB^{\prime}$ can properly explain experimentally observed defect planes that have been incorrectly attributed to Holt-$IDB$ ~\cite{dimitrakopulos_structural_2001,kioseoglou_microstructure_2003}. 
In these experiments, a $(0001)$ planar basal stacking fault (SF) model referred to as the $I_1\;SF$
intersecting with an $IDB^*$, causes the $IDB^*$ to transform to a different configuration.
The defect plane resulting from this transformation was identified as Holt-$IDB$. 
However, the stacking sequence mismatch across the defect plane in question matches the stacking sequence mismatch found in $IDB^{\prime}$ and not Holt-$IDB$.
Given that the $IDB^{\prime}$ model also has a lower domain wall energy compared to Holt-$IDB$, it is more likely to appear in GaN samples.\\

Our paper is organized as follows.
In section~\ref{sec:Notation} we introduce a convenient notation to characterize the different IDB models. 
In section~\ref{sec:Models} we provide illustrated descriptions of each IDB model. 
Details on the creation of supercells containing defects and the computational parameters of our density functional theory (DFT) calculations are presented in section~\ref{sec:Methods}.
We will return to the discussion on the suitability of the $IDB^{\prime}$ model over the Holt-$IDB$ model to describe the defect plane produced by the $IDB^*$\textendash-~$I_1\;SF$ interaction in section~\ref{sec:Discussion}.

\section{\label{sec:Notation}GaN crystal structure and notation for defects}
Before we enter into the description of the creation of models for defects of interest, we briefly review the wurzite crystal structure and describe our notation to clearly identify planar defects.\\

\begin{figure}[tb]
     \centering
     \begin{subfigure}[b]{.23\textwidth  }
         \centering
         \includegraphics[width=\textwidth]{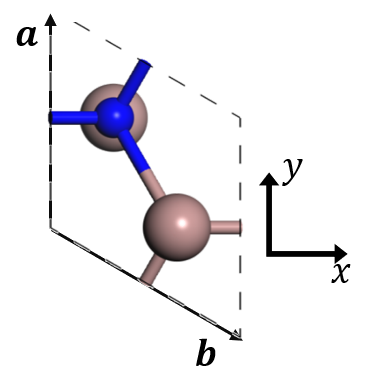}
         \label{fig:prim_unitcell-xy}
         \caption{ }
     \end{subfigure}
     \begin{subfigure}[b]{.23\textwidth}
         \centering
         \includegraphics[width=\textwidth]{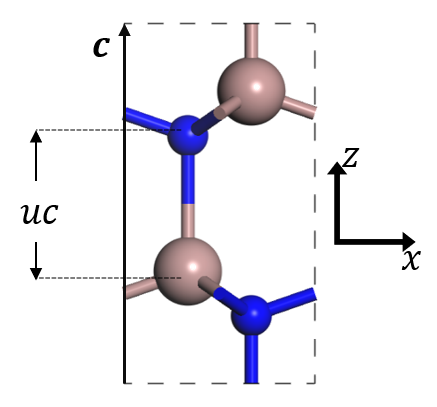}
         \label{fig:prim_unitcell-xz}
         \caption{ }
     \end{subfigure}
     \caption{The primitive unit cell of wurzite GaN viewed onto the (a) $(0001)$ plane and (b) $(1\overline{2}10)$ plane. The Ga (brown) and N (blue) sub-lattices have a relative displacement of $uc$.}
        \label{fig: Primitive unit cell GaN}
\end{figure}

The wurzite crystal structure belongs to the hexagonal crystal system. 
The primitive unit cell of wurzite GaN, shown in figure \ref{fig: Primitive unit cell GaN}, contains a four-atom basis and corresponds to the space group \textbf{P6$_{3}$mc}. 
The lattice parameters obtained by energy minimization are listed in table \ref{tab:opt_lattice params} showing a  $\frac{c}{a}=1.629$ which is $\approx0.2\%$ smaller than the ideal hcp $\frac{c}{a}=\sqrt{\frac{8}{3}}$. 
The wurzite structure has a free parameter, $u\approx\frac{3}{8}$, that determines the bond length of Ga-N dimers parallel to the c axis. \\
\begin{table}[hb]
\caption{\label{tab:opt_lattice params}%
Table of calculated lattice parameters of GaN primitive unit cell compared with experimental results.}
\begin{ruledtabular}
    \begin{tabular}{ccc}
        \textrm{Lattice Parameter}& \textrm{Calculated} &\textrm{Experimental} \\
        \colrule
        \textrm{a} & 3.220 & 3.19 \cite{xia_high-pressure_1993} \\
        \textrm{c} & 5.246 & 5.19 \cite{xia_high-pressure_1993}\\
        \textrm{u} & 0.377 & 0.377 \cite{ueno_stability_1994} 
    \end{tabular}
\end{ruledtabular}
\end{table} 

The wurzite structure can be understood as a hexagonal close packed (hcp) lattice with a dimer basis parallel to the $\mathbf{c}$-axis.
The dimers in the hcp arrangement have an alternating stacking sequence that is denoted by $\dotsi ABAB \dotsi$.
Here $A$ and $B$ refer to the basal plane stacking locations conventionally denoted by ${A,\,B\text{ and }C}$ in hexagonal crystal systems.
If the stacking sequences in the two domains of a planar defect are not identical \emph{and} aligned, we say there is a stacking sequence mismatch across the defect plane. 
Wurzite being non-centrosymmetric allows for spontaneous polarization which manifests in the $\left[0001\right]$ direction~\cite{hellman_polarity_1998,stutzmann_playing_2001}.
For convenience in identifying the polarity of a domain we assign the vector $u\mathbf{c}$ to be the vector pointing from the Ga atom to the N atom in the the Ga-N dimers bonded parallel to the $\mathbf{c}$-axis. 
By convention, if $u\mathbf{c}$ is parallel to $\mathbf{c}$ (i.e. $u>0$), the GaN crystal is referred to as Ga-polar GaN. In the opposite case ($u<0$) we call it N-polar GaN. 
We will use the sign of $u$ to differentiate Ga-polar $(u>0)$ from N-polar $(u<0)$ GaN.
If the polarity in the two domains of a planar defect are in opposing directions, we say there is a polarity reversal across the defect plane.\\
To simplify the description of IDBs and differentiate between possible models, we introduce the following notation. The symbol,
\newcommand{\dplane}[1]{\setlength\ULdepth{-.4ex}\uline{\phantom{#1}}}

\begin{equation}
    \label{eq: def_notation}
    \begin{array}{crccll}
        &\dotsi &\alpha_{I}& \beta_{I}& \dotsi _{sign(u_{I})}&\\
        \multicolumn{5}{c}{\text{\dplane{\kern 25ex}}} & \text{\scriptsize{$\dotsi \{hklm\}$}}\\
        &\dotsi &\alpha_{II}& \beta_{II}& \dotsi _{sign(u_{II})}&
    \end{array}
\end{equation}
specifies the mismatch in polarity and/or stacking sequence across the $\{hklm\}$ plane. For IDBs  $\{hklm\}=\{10\overline{1}0\}$. 
The polarity of each domain is denoted by the $sign(u)$, 
for example, IDBs will correspond to $u_I\neq u_{II}$. 
The stacking sequence of domain I is denoted by $\{\alpha_I,\beta_I\}$, each taking values $\{A,B,C\}$ corresponding to the canonical labeling of the stacking in hexagonal lattices. Notice that, since our stacking elements are dimers, in each domain $\alpha $ should be different from $\beta$ to avoid atomic overlap. 
The ellipses $(\dotsi)$ are used to indicate the periodic repetition of stacking planes; they may be absent in the case of a defect parallel to the basal plane where the sequence terminates.
The stacking sequence mismatch across the defect plane is read as $\alpha_{I}\to\alpha_{II}$ \emph{and} $\beta_{I}\to\beta_{II}$.
Our notation can also be used to denote other planar structural defects such as stacking mismatch boundaries and prismatic stacking faults.\\

In our notation there are multiple ways of specifying the same defect due to the periodicity of the stacking sequence and possible permutations of stacking plane labels.
Consider the two defect notations, 
\newcommand{\defnotation}[6]{
    \begin{array}{crccll}
        &\dotsi &{#1}& {#2} & \dotsi _{({#3})}&\\
        \multicolumn{5}{c}{\text{\dplane{\kern 18ex}}} & \text{\scriptsize{$ \{10\overline{1}0\}$}}\\
        &\dotsi &{#4}& {#5}& \dotsi _{({#6})}&
    \end{array}
}
\begin{align}
    &\defnotation{A}{B}{+}{A}{C}{-}\text{, and}\\[20pt]
    &\defnotation{B}{A}{+}{C}{A}{-}.
\end{align}
Due to the periodicity of the stacking sequences it is clear that both refer to the same type of defect plane. 
Let us consider the following defect notations, 
\begin{align}
    D_1 &= \defnotation{A}{B}{+}{A}{C}{-} \text{, and}\\[20pt]
    D_2 &=\defnotation{A}{B}{+}{C}{B}{-}.
\end{align}
Swapping the stacking labels A and B in $D_2$ results in,
\begin{equation}
    D_2 = \defnotation{B}{A}{+}{C}{A}{-}. 
\end{equation}
This in turn is equivalent $D_1$ as shown previously. 
Two defect notations will represent two \emph{unique} defects if they are not connected by the operation of interchanging stacking site labels \textbf{and/or} they are not equivalent due to periodicity of stacking sequences.
\\
\section{\label{sec:Models}Constructing Defect Models}
\subsection{Inversion Domain Boundaries}
\begin{figure}[t]
    \centering
    \includegraphics[scale=.5]{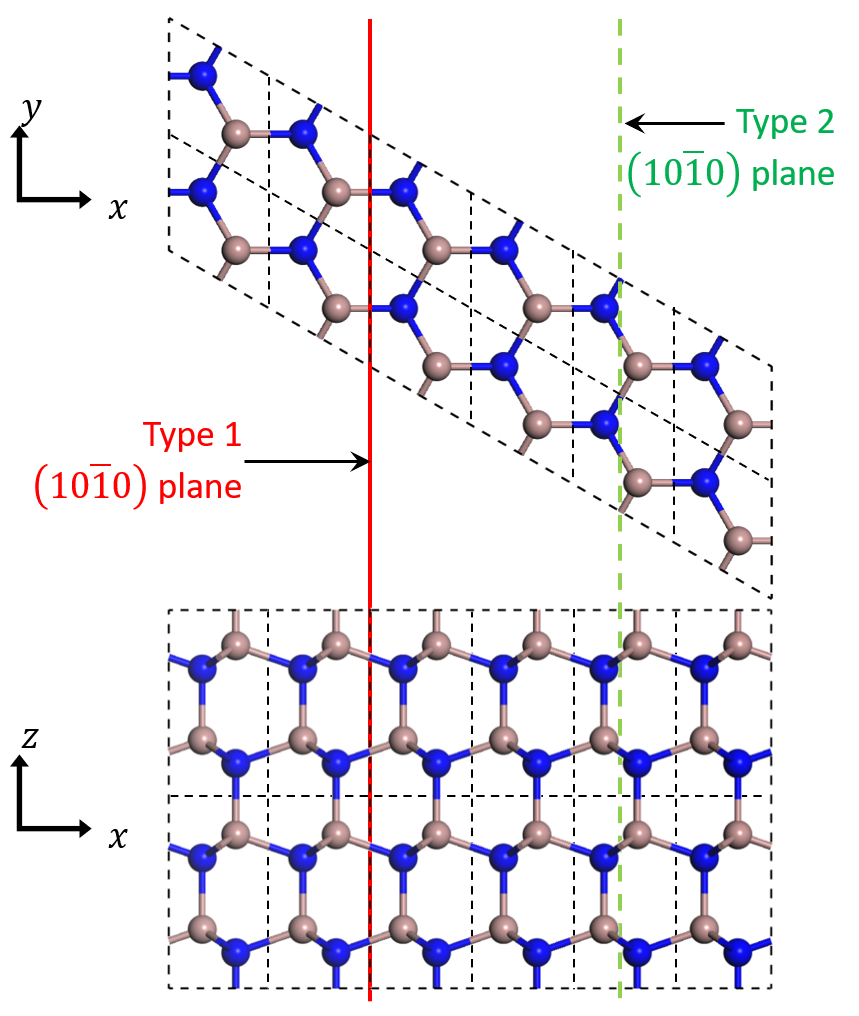}
    \caption{Views of $(0001)$-plane (top) and $(1\overline{2}10)$-plane (bottom)  of wurzite GaN, showing two distinct possible locations for the placement of $\left(10\overline{1}0\right)$ defect planes. The plane indicated by the solid red line corresponds to the type 1 $\left(10\overline{1}0\right)$ plane and the dashed green line corresponds to the type 2 of the $\left(10\overline{1}0\right)$ plane.}
    \label{fig:sdplanes}
\end{figure}

IDBs manifest in the $\{10\overline{1}0\}$ plane.
Planar defects in this plane can occur in one of two types, depending on the location of the plane and the proximity of neighboring atoms to the defect plane. 
As depicted in Fig.~\ref{fig:sdplanes}, there are sparse $\{10\overline{1}0\}$ planes, conventionally called type 1, and dense $\{10\overline{1}0\}$ planes, called type 2.
Our calculations and previous works ~\cite{kioseoglou_interatomic_2008,zhang_thermodynamics_2018} indicate that for all studied cases of $\{10\overline{1}0\}$ planar defects the type 1 defects are energetically favorable compared to the type 2 defects. 
Thus, in the rest of this document, unless explicitly stated otherwise any $\{10\overline{1}0\}$ planar defect we discuss or illustrate will be of type 1.\\ 

IDBs are $\{10\overline{1}0\}$ planar defects that separate two domains where one domain is Ga-polar and the other in N-polar. 
Some of them may also have a mismatch in crystal stacking sequence. 
In our notation (\cref{eq: def_notation}), a generic IDB can be denoted by,
\renewcommand{\defnotation}[3]{
    \begin{array}{crccll}
        &\dotsi &{A}& {B} & \dotsi _{({+})}&\\
        \multicolumn{5}{c}{\text{\dplane{\kern 18ex}}} & \text{\scriptsize{$ \{10\overline{1}0\}$}}\\
        &\dotsi &{#1}& {#2}& \dotsi _{({#3})}&
    \end{array}
}
\begin{equation}
    \label{eq:generic idb notation}
    \defnotation{\alpha}{\beta}{-},
\end{equation}
where by convention we choose $\dotsi AB\dotsi_+$ for the first domain.
Considering that $\alpha$ and $\beta$ can take three values that are different, the total number of possible IDBs would be apparently six. However, there are only four \emph{unique} configurations.
Two of these correspond to the Holt-$IDB$ and $IDB^*$ models described in the Introduction. 
The other two will be labeled $IDB^{\prime}$ and $IDB^{\prime\prime}$. 
To the best of our knowledge, $IDB^{\prime}$ and $IDB^{\prime\prime}$ have not been discussed or evaluated for energetic viability before.
Here a $\{10\overline{1}0\}$ defect model is considered to be energetically viable if the domain wall energy is less than twice the $\{10\overline{1}0\}$ surface energy $(\Sigma_{\{10\overline{1}0\}})$.\\

\begin{figure*}[t]
     \centering
     \begin{subfigure}[t]{.24\textwidth}
         \centering
         \includegraphics[width=\textwidth]{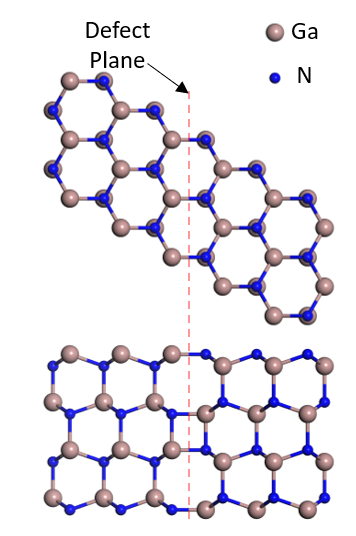}
         \caption{$IDB^*$\\
         $\Gamma_{IDB^*} = 17.7\;\text{m$\,$eV/\AA}^2$}
         \label{fig:IDBs model}
     \end{subfigure}
     \hfill
     \begin{subfigure}[t]{.24\textwidth}
         \centering
         \includegraphics[width=\textwidth]{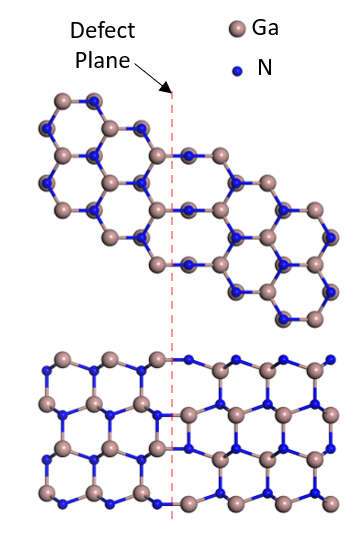}
         \caption{$IDB^{\prime}$\\
         $\Gamma_{IDB^{\prime}} = 131.6\;\text{m$\,$eV/\AA}^2$}
         \label{fig:IDBp model}
     \end{subfigure}
     \hfill
     \begin{subfigure}[t]{.24\textwidth}
         \centering
         \includegraphics[width=\textwidth]{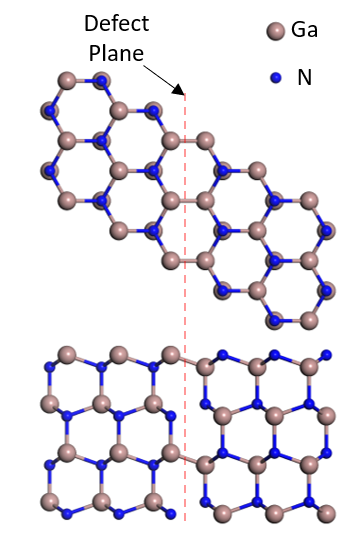}
         \caption{Holt-IDB\\
         $\Gamma_{Holt-IDB} = 195.2\;\text{m$\,$eV/\AA}^2$}
         \label{fig:IDB-Holt model}
     \end{subfigure}
     \hfill
     \begin{subfigure}[t]{.24\textwidth}
         \centering
         \includegraphics[width=\textwidth]{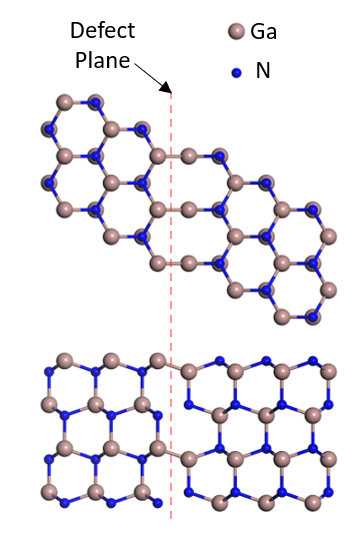}
         \caption{$IDB^{\prime\prime}$\\
         $\Gamma_{IDB^{\prime \prime}} = 246.3\;\text{m$\,$eV/\AA}^2$}
         \label{fig:IDBpp model}
     \end{subfigure}
        \caption{The views onto $\left(0001\right)$ plane (top) and $\left(1\overline{2}10\right)$ plane (bottom) are shown for $IDB^*$, $IDB^{\prime}$, Holt-$IDB$ and $IDB^{\prime\prime}$ defect models. The energetically favorable type 1 defects are shown.}
        \label{fig:IDB models}
\end{figure*}

The four possible IDB models are shown in \cref{fig:IDB models} in order of increasing domain wall energy$(\Gamma)$ and we describe them in more detail here.
\begin{enumerate}[leftmargin=*,labelsep=0pt,align=left,label={\roman*.}]
    \item \textbf{Holt-$IDB$}\\
    \begin{equation}
        \label{not:Holt}
        \defnotation{A}{B}{-}
    \end{equation}
    Holt-$IDB$ is produced by the interchanging of species sub-lattices in one of the two adjacent domains separated by a $\{10\overline{1}0\}$ plane. 
    Across the defect plane the stacking planes are undisturbed, i.e. stacking planes $A\to A$ \emph{and} $B\to B$. 
    Bonds that cross defect plane will be `wrong bonds' between atoms of the same species as shown in \cref{fig:IDB-Holt model}.\\
    
    \item \textbf{$IDB^*$}\\
    \begin{equation}
        \label{not:IDBs}
        \defnotation{B}{A}{-}
    \end{equation}
    $IDB^*$ is produced introducing a $\tfrac{1}{2}\mathbf{c}$ relative translation between the two polarity reversed regions in the Holt-$IDB$ model. 
    This translation introduces a mismatch in stacking sequence across the defect plane, i.e. $A\to B$ \emph{and} $B\to A$. 
    Bonds across the defect plane are not `wrong bonds'.
    Along \textbf{c}, the  bonds alternately form co-planar four-membered rings and non-co-planar eight-membered rings as depicted in \cref{fig:IDBs model}.\\

    \item \textbf{$IDB^{\prime}$}\\
    \begin{equation}
        \label{not:IDBp}
        \defnotation{C}{A}{-}
    \end{equation}
    From the $IDB^*$ model, altering one of the stacking planes in one of the regions to the unoccupied stacking plane location produces the $IDB^{\prime}$ defect model. 
    The stacking sequence mismatch in $IDB^{\prime}$ can be, $A\to B$ \emph{and} $B\to C$ or $A\to C$ \emph{and} $B\to A$. 
    The change of stacking to the unoccupied stacking plane location ($C$ here) introduces a net translation of atoms perpendicular to the defect plane. 
    If this translation is away from the defect plane, type 1 $IDB^{\prime}$ is formed. 
    If the translation is toward the defect plane, type 2 $IDB^{\prime}$ is formed. 
    Similar to $IDB^*$ bonds across the defect plane form four-membered and eight-membered rings alternately along $\mathbf{c}$. 
    However, the eight-membered rings are asymmetrical. 
    Additionally here, the bonds across the defect plane are not `wrong bonds' as seen in \cref{fig:IDBp model}.\\
    
    \item \textbf{$IDB^{\prime\prime}$}\\
    \begin{equation}
        \label{not: IDBpp}
        \defnotation{A}{C}{-}
    \end{equation}
    From the Holt-$IDB$ model, altering one of the stacking planes in one of the regions to the unoccupied stacking plane location produces the $IDB^{\prime\prime}$ defect model. 
    The stacking sequence mismatch in $IDB^{\prime\prime}$ can be, $A\to A$ \emph{and} $B\to C$ or $A\to C$ \emph{and} $B\to B$. 
    Here also, the change of stacking to the unoccupied stacking plane location ($C$ here) introduces a net translation of atoms perpendicular to the defect plane. 
    And the nature of this translation (away from or toward the defect plane) will determine whether type 1 or type 2 $IDB^{\prime\prime}$ is formed. 
    Also here, the bonds across the defect plane are `wrong bonds' as seen in \cref{fig:IDBpp model}.
\end{enumerate}

The two existing IDB models can be distinguished based on the stacking sequence mismatch (or lack thereof) across the defect plane. 
This distinction between relative stacking sequences of the two domains, has been used on TEM images of GaN samples to experimentally confirm the existence of $IDB^*$ and Holt-$IDB$~\cite{cherns_determination_1998,potin_1010_1999,potin_evidence_1999}. 
In section~\ref{sec:Discussion} we will use the same argument to show that our model, $IDB^{\prime}$, is more suitable to describe the defect plane resulting from the $IDB^*$~\textendash~$I_1\;SF$ interaction.
\\

In addition to these IDB models there is another historical model, proposed by Austerman and Gehman~\cite{austerman_inversion_1966}. 
The Austerman model is equivalent to translating the polarity reversed region of the $IDB^*$ model by $\sim-\tfrac{1}{8}\mathbf{c}$ such that the Nitrogen sub-lattice is uninterrupted by the defect plane. 
The motivation for this model was to avoid distorting the sub-lattice corresponding the species with the larger ions. 
In Beryllium Oxide, for which Austerman's model was proposed, the anions are larger. 
In the case GaN the reverse is true and therefore we can conceive a Austerman$^{*}$ model where we do a $\sim +\tfrac{1}{8}\mathbf{c}$ shift of the polarity reversed region to make the Ga atoms co-planar in $\left(0001\right)$ planes. 
Additionally we can then play the game of changing the relative stacking sequences of the two domains of Austerman and Austerman$^*$ models to create a total of eight Austerman based inversion domain boundary models. 
However, all these models were unstable and found to relax to one of the previously listed four models $IDB^*$, $IDB^{\prime}$, Holt-$IDB$ or $IDB^{\prime\prime}$.\\

\subsection{Stacking Mismatch Boundaries}
A stacking mismatch boundary is a $\{10\overline{1}0\}$ plane separating two domains that have a stacking sequence mismatch and the \emph{same} polarity. 
A generic stacking mismatch boundary model can be denoted by, 
\begin{equation}
    \label{eq: generic SMB notation}
    \defnotation{\alpha}{\beta}{+}.
\end{equation}
Our methodology of enumerating stacking sequence mismatches across the defect plane can also be used to create four possible models for stacking mismatch boundaries. 
One of these is the accepted model for stacking mismatch boundaries in the literature. 
Of the additional stacking mismatch boundary models we produce, one is  trivial (no stacking mismatch) and the other two fail the test for energetic viability. 
We will refer the sole viable model for stacking mismatch boundaries as $SMB$ since there is no ambiguity. 
In our notation $SMB$ is represented by, \begin{equation}
    \label{not: SMB}
    \defnotation{A}{C}{+}
\end{equation} 
The stacking sequence mismatch across the defect plane is the same as the $IDB^{\prime\prime}$ model, i.e. $A\to A$ \emph{and} $B\to C$ or $A\to C$ \emph{and} $B\to B$. 
Similar to the $IDB^{\prime\prime}$ (and $IDB^{\prime}$) the change of stacking plane to the third possibility introduces a net translation of atoms perpendicular to the defect plane and the direction of this translation determines whether a type I or type 2 defect model is produced.
In section~\ref{sec:Methods} we will discuss the necessity of devising an improved method to estimate domain wall energies of defect models that involve all three stacking plane possibilities. 
This method improves the estimation of the domain wall energy of the $SMB$ model.
As such, we will include the $\Gamma_{SMB}$ in our final tabulation of results in section~\ref{sec:Discussion}.

\subsection{Basal Stacking Faults}
\label{subsec: SF models}
  Since we are interested in the transformation of IDBs via interaction with basal stacking faults(SF),
 we provide a brief review of SFs here.
 SFs occur when the stacking sequence of the GaN crystal is disrupted at a  basal plane, i.e. a $(0001)$ plane. 
 Alternately, SFs can be considered as wurzite-sphalerite transformation for a few atomic layers. 
 Sphalerite also has dimer occupation of stacking planes similar to wurzite. 
 In the sphalerite structure the stacking sequence has a periodicity of three stacking planes $(\dotsc ABCABC \dotsc)$, unlike the periodic repetition of two stacking planes $(\dotsc ABAB \dotsc)$ of the wurzite structure. 
 We are aware of three models for SFs described in the literature; they are referred to as $I_1\;SF$, $I_2\;SF$, and $E\:SF$. 
 The SF models respectively correspond to three, four, and five basal stacking planes in the sphalerite region. 
 Our calculations for domain wall energies yield $\Gamma_{I_1\;SF}=1.1\;m\,eV/\mathring{A}^2$, $\Gamma_{I_2\;SF}=2.5\;m\,eV/\mathring{A}^2$, and $\Gamma_{E\:SF}=3.9\;m\,eV/\mathring{A}^2$.
 These values are in agreement with previously reported  values~\cite{wright_basal-plane_1997,stampfl_energetics_1998}.\\
 
 The different SF models can be represented by specifying the stacking sequence in the vicinity of the defect. For example, $I_1\;SF$ can be denoted by the sequence $\dotsc AB\mathbf{ABC}BC\dotsc$. 
 Here the stacking planes given in \textbf{bold} correspond to the sphalerite region. 
 SF models have varying thickness depending on the thickness of the sphalerite region, the middle of the sphalerite region is considered to be SF location.  
 If the SF model contains an odd number of stacking planes in the sphalerite region ($I_1\;SF$ and $E\:SF$), the defect plane location coincides with the central  stacking plane in the sphalerite region. 
 We will use a vertical bar $(|)$ to indicate the location of the stacking fault. 
 As such $I_1\;SF$ is represented by, 
 \begin{equation}
     \label{not:I_1 SF}
     \dotsi AB\mathbf{A}\mathrlap{\;|}\mathbf{BC}BC \dotsc.
 \end{equation}
 If the there are an even number of stacking planes in the sphalerite region the fault location is in between two stacking planes. 
 So, we denote $I_2\;SF$ by, 
 \begin{equation}
    \label{not:I_2 SF}
    \dotsc AB \mathbf{AB|CA} CA\dotsc.
 \end{equation}
 $E\:SF$ is equivalent to an $I_2\;SF$ being immediately followed by an $I_1\;SF$ as can be seen from the stacking sequence corresponding to $E_SF$,
 \begin{equation}
     \label{not:E SF}
     \cdots AB
     \underbrace{\vphantom{\frac{A}{B}}
     \rlap{
     $\overbrace{
     \vphantom{
        \begin{smallmatrix}M\\M\\M
        \end{smallmatrix}}
     \phantom{\mathbf{AAA}}}^{I_2\;SF}
     $}
     \mathbf{AB\mathrlap{\;|}{C}AB}}_{E\:SF}
     \llap{$\overbrace{
     \vphantom{\begin{matrix}M\\M\\M\\M\end{matrix}}
     \phantom{\mathbf{BCll}}}^{I_1\;SF}$}
     AB\cdots 
 \end{equation}.

\section{\label{sec:Methods}Methods}
\subsection{\label{subsec:calculate Gamma}Calculating Domain Wall Energies of Defects}
We define the domain wall energy of a defect as,
\begin{equation}
    \label{eq:Gamma def}
    \Gamma_{\text{defect}}=\frac{E_{\text{supercell with defects}}-E_{\text{reference}}}{A_{\text{defect}}}
\end{equation}
where, $E_{\text{supercell with defects}}$ is the total energy of a supercell containing the defect, $A_\text{{defect}}$ is the area of the defect in the defect supercell, and $E_{\text{reference}}$ is the total energy of a reference supercell that contains an equal number of atoms as the defect supercell. \\

The sizes of the supercells we create will be denoted by $n_{a}\times n_{b}\times n_{c}$, where $n_a$, $n_b$ and $n_c$ are the number of repeated primitive unit cells of wurzite GaN along the $\mathbf{a}$, $\mathbf{b}$ and $\mathbf{c}$ directions. 
When creating defect supercells, the periodicity of the supercells dictate that at least two defect planes are present. 
We attempt to minimize interaction of defect planes by creating supercells with six primitive unit cells between defect planes. 
We performed calculations using larger defect supercells, with larger defect plane separation (upto nine primitive unit cells between defect planes).
The results indicate that domain wall energies are converged with respect to defect plane separation for supercells six primitive unit cell defect plane separation. 
This requirement corresponds to the $\{10\overline{1}0\}$ defect planes having a perpendicular separation of $6a\cos{(\pi/6)}$.\\

\begin{figure}[bth]{}
    \centering
    \includegraphics[scale=.7]{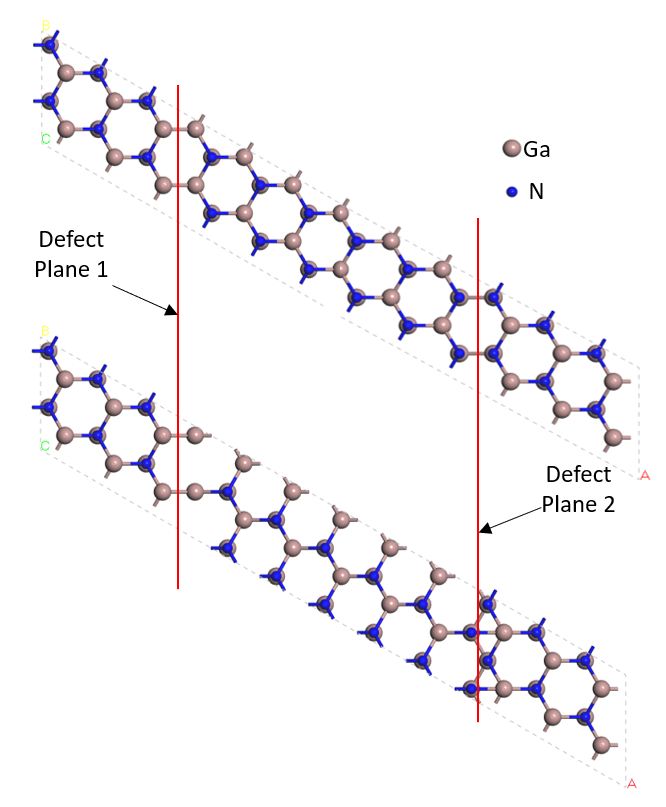}
    \caption{$\left(0001\right)$ planar views of $12\times2\times1$ supercells containing, two Holt-$IDB$s (top) and two $IDB^{\prime\prime}$s (bottom). In both the cases the $\left(01\overline{1}0\right)$ planar defects are separated by $6a\cos(30^{\circ})$.}
    \label{fig:ad vs bc defects}
\end{figure}
The definition of $\Gamma_{\text{defect}}$ in \cref{eq:Gamma def} assumes the defect supercell under consideration contains only one type of defect. 
However, for $\{10\overline{1}0\}$ planar defects that include all three stacking locations ($IDB^{\prime}$, $IDB^{\prime\prime}$, and $SMB$), creating defect supercells containing two defect planes of the same configuration is impossible.  
Occupying the third stacking location across the defect plane results in a net translation of atoms perpendicular to the defect plane. 
This results in one of the planes being the sparse form and the other plane being the dense form of the same defect model.
To illustrate this complication, we show $\left(0001\right)$ planar views of $12\times 2\times 2$ supercells containing Holt-$IDB$ and $IDB^{\prime\prime}$ in \cref{fig:ad vs bc defects}. 
Defect planes 1 and 2 for the Holt-$IDB$ containing supercell are identical with a $\frac{1}{2}\mathbf{c}$ relative shift between them. 
Thus, the supercell containing Holt-$IDB$ shown in \cref{fig:ad vs bc defects} can be used with equation \eqref{eq:Gamma def}. 
However, the defect planes 1 and 2 in the $IDB^{\prime\prime}$ containing supercell are not the same type and in fact respectively correspond to the type 1 and type 2 forms of the $IDB^{\prime\prime}$ defect model. 
Using this supercell with \cref{eq:Gamma def} will give the average domain wall energy of the type 1 $IDB^{\prime\prime}$ and type 2 $IDB^{\prime\prime}$ models. 
A previous report~\cite{northrup_inversion_1996} on the domain wall energy of $SMB$ seems to report this average value.\\

For the $IDB^{\prime}$, $IDB^{\prime\prime}$, and $SMB$ defects, it is possible to create supercells with only one version of defect \emph{if} six defect planes are included. 
However, since each defect plane involves a translation of atoms perpendicular to the plane the volume per atom in the defect supercell will be different from the volume per atom of pristine GaN. 
In \cref{fig:ad vs bc defects} we see that a single defect plane translates a GaN dimer $\frac{1}{3}\mathbf{a}$ away from (towards) the boundary for the type 1 (type 2) defect. 
Since a GaN primitive unit cell has two GaN dimers, over six defect planes, both GaN dimers are translated by $\mathbf{a}$. 
Therefore, for type 1 (type 2) of each model we create a supercell that occupies the volume of $37(35)\times1\times1$  primitive unit cells and contains the same number of atoms in 36 primitive unit cells. 
For the reference supercell we create a $36\times1\times1$ pristine supercell so that the total number of atoms are the same.

\subsection{\label{subsec:computational details}Computational details}
We employed Density Functional Theory (DFT) calculations using the Vienna Ab-initio Simulation Package (VASP)~\cite{kresse_efficient_1996,kresse_efficiency_1996}. 
The exchange correlation interactions were treated using the generalized gradient approximation (GGA) with the Perdew-Burke-Ernzerhof (PBE) functional~\cite{perdew_generalized_1996}. 
The projector augmented wave(PAW) potentials~\cite{blochl_projector_1994,kresse_ultrasoft_1999} recommended and provided by  the VASP code were used to treat core states. 
We use pseudopotentials with valence electron configurations 3d$^{10}$4s$^2$4p$^1$ for Ga and 2s$^2$2p$^3$ for N. 
The cutoff energy and k-point mesh were chosen such that the calculated total energies of the supercells of interest were accurate to 1 $m\,eV$. 
For this, we chose a cutoff energy of 900 $eV$.
The k-point mesh used to sample the reciprocal space was a Monkhorst-Pack grids~\cite{monkhorst_special_1976} centered at the Gamma point. 
The k-point meshes were automatically generated using the VASP scheme to generate k-point meshes by specifying a length parameter $(R_k)$.
This grid scheme produces, $N_i=int(max(1, R_k\times|\vec{b}_i| + 0.5))$ divisions along each reciprocal lattice vector ($\{\vec{b}_i\}$). 
\cref{tab:kmesh} lists the resulting k-mesh grids used depending on the supercell size.\\

\begin{table}[bth]
    \caption{\label{tab:kmesh}%
    Table of size of supercells considered and corresponding k-point grids used.}
    \begin{ruledtabular}
        \begin{tabular}{cc}
            \textrm{Supercell Size}& \textrm{Monkhorst-Pack}
            \\
            \textrm{($n_{a}\times n_{b}\times n_{c}$)} & \textrm{k-point grid}\\
            \colrule
            $1\times1\times1$ & $9\times9\times5$ \\
            $12\times1\times1$ & $1\times9\times5$\\
            $\left(36\pm n\right)\times1\times1$\footnote{$n\in\{0,1\}$} & $1\times9\times5$ \\
            $1\times1\times12$ & $9\times9\times1$\\
        \end{tabular}
    \end{ruledtabular}
\end{table}

\section{\label{sec:Discussion}Discussion}
The domain wall energies of the $\{10\overline{1}0\}$ planar-defect models discussed previously are listed in \cref{tab:gamma_values}. 
Each model is evaluated for energetic viability by the condition  ${\Gamma\,<\,2\Sigma_{\{10\overline{1}0\}}}$, according to our calculations ${\Sigma_{\{10\overline{1}0\}}=98.7\;m\,eV/\text{\AA}^2}$. 
\\
\begin{table}[htb]
    \caption{\label{tab:gamma_values}%
    Domain wall energies of type 1 and type 2 defect models for IDB models and $SMB$. For type 1 models if available literature values are listed.}
    \begin{ruledtabular}
        \begin{tabular}{cccc}
            \textrm{ }& \multicolumn{3}{c}{Domain Wall Energy, $\Gamma$  (m$\,$eV/\AA$^2$)}\\
            \textrm{Defect}&\multicolumn{3}{c}{\hrulefill}\\
            \textrm{Model} & Type 1  & \textrm{Literature value}&Type 2 \\
            \colrule
            \textrm{Holt-$IDB$}&195.2 & 167~\cite{northrup_inversion_1996}&253.2\footnotemark[1]\\
           \textrm{$IDB^*$} & 17.7 & 25~\cite{northrup_inversion_1996}&44.7\\
            \textrm{$IDB^{\prime}$}&131.6&-&172.4\\
            \textrm{$IDB^{\prime\prime}$} & 246.3\footnotemark[1] & - & 392.9\footnotemark[1]\\
            \colrule
            \textrm{$SMB$} & 93.9 & 105~\cite{northrup_inversion_1996} & 171.8 \\
        \end{tabular}
    \end{ruledtabular}
    \footnotetext[1]{These $\Gamma$ values do not satisfy the energetic viability condition, $\Gamma < 2\Sigma_{\{10\overline{1}0\}}$.}
\end{table}

Of the IDB models, $IDB^{\prime\prime}$ and type 2 Holt-$IDB$ fail the viability test and can be discarded. 
$IDB^*$, proposed by Northrup, type 1 and type 2, has significantly lower domain wall energy compared to the other models and is confirmed to be the preferred configuration for IDBs. 
The $IDB^{\prime}$, although energetically unfavorable compared $IDB^*$, is physically viable and may occur in GaN samples. 
During epitaxial growth of GaN samples, in some cases, the formation of $IDB^{\prime}$ may be kinetically favored.
\\

One possible example of $IDB^{\prime}$ forming can be found in the work of Kong et al.~\cite{kong_titanium_2016}. This work studies the formation of inversion domains in GaN nano-columns grown using MBE on Titatinum(Ti) masked GaN substrates. 
The authors attribute the formation of inversion domains to a residual $(0001)$ Ti monolayer. 
Various configurations of the interfacial TiN monolayer and the subsequent growth of the  polarity inverted region are studied. 
In the lowest energy Ti monolayer configuration (see Fig. 6 of \cite{kong_titanium_2016}), the stacking sequence mismatch between the inversion domain and the nano-column matrix corresponds to the stacking sequence mismatch of $IDB^{\prime}$. Therefore we expect $IDB^{\prime}$ defects to be found at the boundary of the Ti monolayer.\\
\begin{figure*}[t]
    \centering
    \begin{subfigure}[b]{.32\textwidth}
        \centering
        \includegraphics[width=\textwidth]{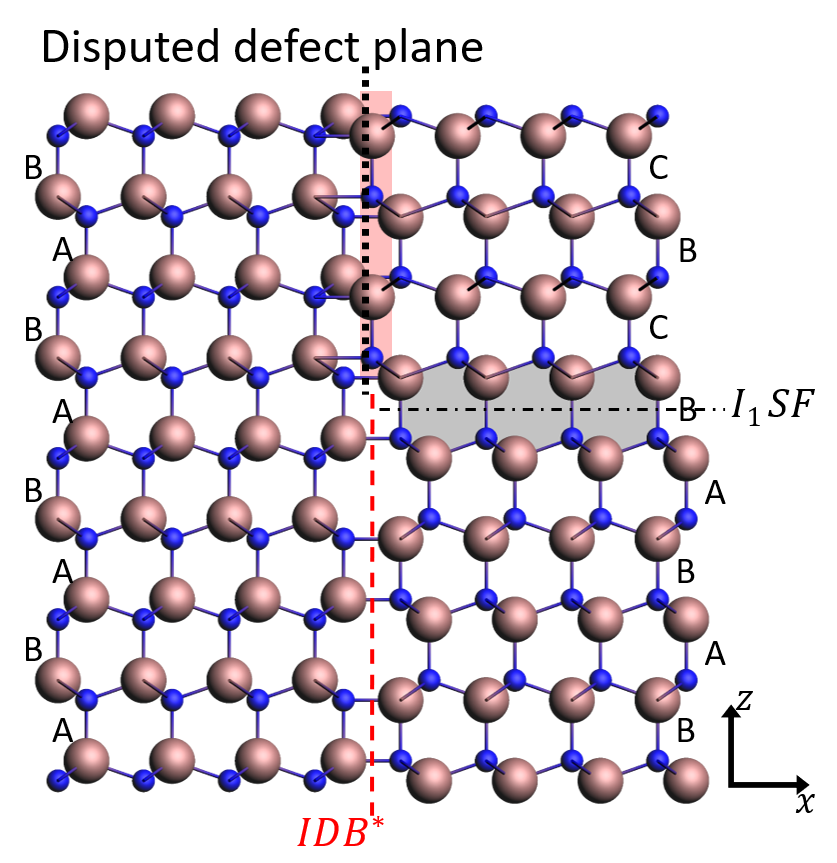}
        \caption{ }
        \label{fig:IDB*-I1 kio model}
    \end{subfigure}
    \hfill
    \begin{subfigure}[b]{.32\textwidth}
        \centering
        \includegraphics[width=\textwidth]{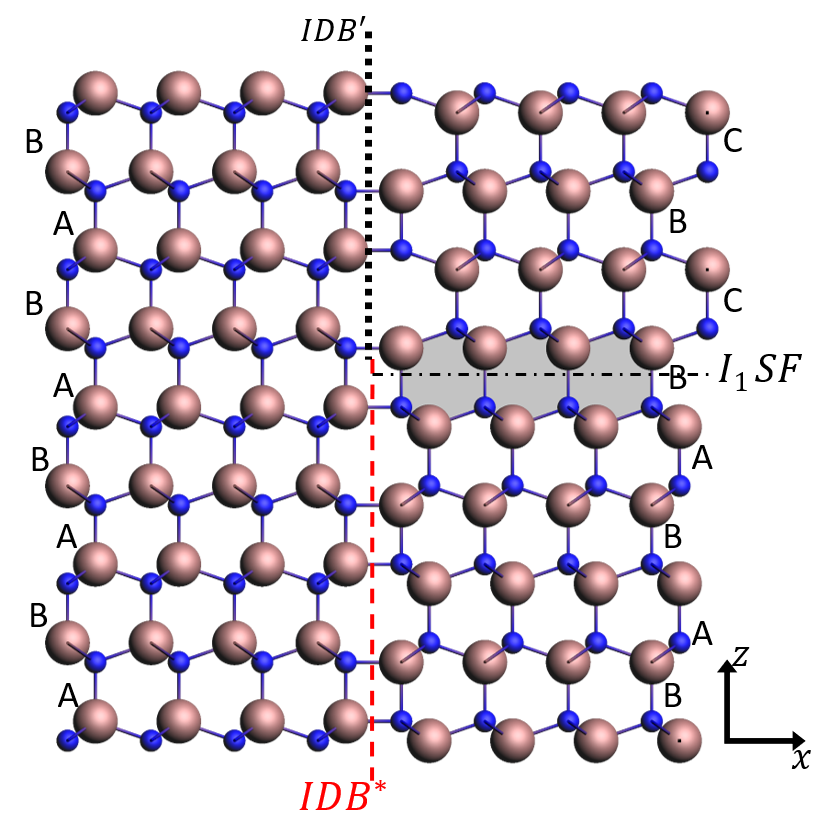}
        \caption{ }
        \label{fig:IDB*-I1 our model}
    \end{subfigure}
    \hfill
    \begin{subfigure}[b]{.32\textwidth}
         \centering
         \includegraphics[width=\textwidth]{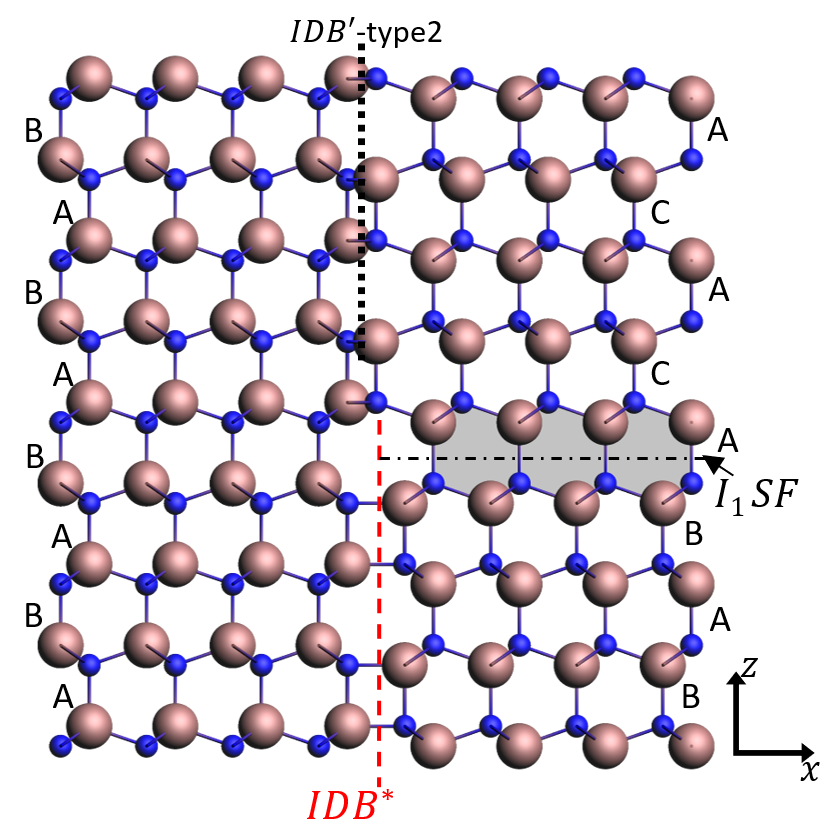}
         \caption{ }
         \label{fig:IDB*-I1 our model alternate}
    \end{subfigure}
    \caption{(a) Previously proposed atomic model for the $IDB^*$\textendash$I_1\;SF$ interaction, (b) Atomic model resulting from removal of spurious atoms (highlighted by the region shaded red in (a)) in the previously proposed model, and (c) $IDB^*$\textendash$I_1\;SF$ interaction with $I_1\;SF$ shifted by $\frac{1}{2}\mathbf{c}$ such that the SF plane is terminated by an eight-membered ring of $IDB^*$.}
        \label{fig:IDB*-I1 models}
\end{figure*}

The most compelling argument for the possible existence of $IDB^{\prime}$ is found in the works by Dimitrakopulos et al.~\cite{dimitrakopulos_structural_2001} and Kioseoglou et al.~\cite{kioseoglou_microstructure_2003}. 
These authors study the transformation of $IDB^*$ via an interaction with an $I_1$ Stacking Fault ($I_1\;SF$). 
Dimitrakopulos et al.\ and Kieseoglou et al.\ identify the IDB resulting from the $IDB^*$\textendash~$I_1\;SF$ interaction as the Holt-$IDB$ model. 
They provide the atomic model shown in \cref{fig:IDB*-I1 kio model} (Fig. 11c in \cite{dimitrakopulos_structural_2001} and Fig. 3b in \cite{kioseoglou_microstructure_2003}) to justify this claim.
We can extend our notation to analyze the combined defects shown in \cref{fig:IDB*-I1 kio model,fig:IDB*-I1 our model}. 
The notation for this atomic model will be,
\begin{equation}
    \label{not:IDB*-I_1}
    \begin{array}{rrccccccccll}
         &\dotsi&A&B&A&B&A&B&A&B&\dotsi_{(+)}\\
         \multicolumn{11}{c}{\text{\dplane{\kern 40ex}}}&\text{\scriptsize{$ \{10\overline{1}0\}$}}\\
         &\dotsi&B&A&B&\mathbf{A}&\mathrlap{\text{ }|}{\mathbf{B}}&\mathbf{C}&B&C&\dotsi_{(-)} 
    \end{array}.
\end{equation}
Note, the orientation of the image is rotated $90^{\circ}$ clockwise in the notation. 
The stacking sequence mismatch across the $\{10\overline{1}0\}$ plane, to the right of the SF plane in \cref{not:IDB*-I_1} (above the SF plane in \cref{fig:IDB*-I1 kio model,fig:IDB*-I1 our model alternate}), is in agreement with that of $IDB^{\prime}$ ($A\to B$ and $B\to C$) and not that of the Holt-$IDB$ ($A\to A$ and $B\to B$). 
Consequently, we propose that the correct interpretation of this image is that the SF transforms a $IDB^*$ defect into an $IDB^{\prime}$.\\

We assume that the erroneous interpretation derives from the fact that the authors did not know about the  $IDB^{\prime}$ model at the time of writing their work. 
Moreover, the atomic model proposed by the authors (\cref{fig:IDB*-I1 kio model}), has a high density of atoms at the interface. 
We have identified the spurious atoms (enclosed in the shaded region of \cref{fig:IDB*-I1 kio model}) that cause this increase in density. 
Upon removing these extra atoms, the model obtained is that of \cref{fig:IDB*-I1 our model} which corresponds exactly to the proposed $IDB^{\prime}$  structure (\cref{fig:IDBp model}). 
We also notice that if the SF is shifted by $\tfrac{1}{2}\mathbf{c}$, the resulting defect would be an $IDB^{\prime}$-type 2. 
In this scenario (\cref{fig:IDB*-I1 our model alternate}), the SF is terminated by an 8 member ring, as opposed to the four member ring of \cref{fig:IDB*-I1 our model}.
This is a feature that potentially can be experimentally discerned.\\ 

The $IDB^*$\textendash-~$I_1\;SF$ interaction was proposed as a possible explanation for experimental observations of Holt-$IDB$ in GaN samples~\cite{dimitrakopulos_structural_2001,kioseoglou_interatomic_2008}. 
As we have shown the  $IDB^*$\textendash-~$I_1\;SF$ interaction results in $IDB^{\prime}$ and thus cannot explain the observations of Holt-$IDB$s. 
Nevertheless, we can look at the other SF models described in section~\ref{sec:Models} and their interaction with $IDB^*$ to explain the formation of Holt-$IDB$s. 
We analyze the $IDB^*$\textendash-~SF~model interactions with the aid of our notation. 
Note the notation does not differentiate between atomic models that have different SF locations and/or include spurious atoms (\cref{not:IDB*-I_1} represents \cref{fig:IDB*-I1 kio model,fig:IDB*-I1 our model,fig:IDB*-I1 our model alternate}); these considerations must be treated according to our discussion in the preceding paragraph. \\

The $IDB^*$\textendash-~$I_2\;SF$ interaction can be represented in our notation as,
\begin{equation}
    \label{not:IDB*-I_2}
    \begin{array}{rrcccccccccll}
         &\dotsi&A&B&A&B&A&B&A&B&A&\dotsi_{(+)}\\
         \multicolumn{12}{c}{\text{\dplane{\kern 40ex}}}&\text{\scriptsize{$ \{10\overline{1}0\}$}}\\
         &\dotsi&B&A&B&\mathbf{A}&\mathbf{B\rlap{\kern .5ex$\mathbf{|}$}}&\mathbf{C}&\mathbf{A}&C&A&\dotsi_{(-)} 
    \end{array}.    
\end{equation}
Now the stacking sequence mismatch of the transformed $\{10\overline{1}0\}$ plane (right of SF plane in \cref{not:IDB*-I_2}) is the same as $IDB^{\prime\prime}$ ($A\to A$ and $B\to C$). 
However, as we have stated previously $IDB^{\prime\prime}$ is energetically not viable and is unlikely to appear in GaN samples. 
This impasse can be resolved in one of two ways. 
The first would be to remove the $I_2\;SF$ by introducing a $1/3<1\overline{1}00>$ basal shear at the $I_2\;SF$ plane. 
This is feasible since $I_2\;SF$ is the only SF model that can be formed (or removed) through strain relaxation. 
In this scenario removing the $I_2\;SF$ results in no transformation of the $IDB^*$. 
The second way of avoiding forming $IDB^{\prime\prime}$ is to include one more stacking plane in the sphalerite sequence, i.e. $I_2\;SF$ becomes $E\:SF$. 
This scenario can be denoted by,
\begin{equation}
    \label{not:IDB*-E}
    \begin{array}{rrccccccccccll}
         &\dotsi&A&B&A&B&A&B&A&B&A&B&\dotsi_{(+)}\\
         \multicolumn{13}{c}{\text{\dplane{\kern 45ex}}}&\text{\scriptsize{$ \{10\overline{1}0\}$}}\\
         &\dotsi&B&A&B&\mathbf{A}&\mathbf{B}&\mathrlap{\text{ }\mathbf{|}}{\mathbf{C}}&\mathbf{A}&\mathbf{B}&A&B&\dotsi_{(-)} 
    \end{array}.    
\end{equation}
Now the stacking sequence mismatch of the transformed$\{10\overline{1}0\}$ defect is consinstent with Holt-$IDB$ ($A\to A$ and $B\to B$). 
Hence, it is possible for $IDB^*$ to transform to the Holt-$IDB$ only when interacting with an $E\:SF$.\\

The work by Lan\c{c}on et al.~\cite{lancon_towards_2018} studies picometer-scale inter-domain shifts parallel to the polarization axis, in supercells containing $IDB^*$. 
In the $IDB^*$ structure, atomic layers in basal planes are occupied by opposing species (i.e. Ga $\to$ N or N $\to$ Ga) across the defect plane. 
Lan\c{c}on et al.\ define the inter-domain shift ($\delta z$) as, the difference the between the $\mathbf{c}$-axis coordinate of a Ga atom basal plane in the Ga-polar domain and the corresponding N atomic plane in the N-polar domain. 
In the $IDB^*$ structure proposed by Northrup et al.~\cite{northrup_inversion_1996}, prior to geometry optimization, $\delta z_{initial}=0$. 
Lan\c{c}on et al.\ note that when an initial inter-domain shift $(\delta z_{initial} > ~ 3.1\,pm)$ is introduced, the structure relaxes to a configuration with a final inter-domain shift $\delta z= 8\;pm$. 
If no initial inter-domain shift is introduced the relaxed configuration has an inter-domain shift $\delta z=-1\;pm$. 
Further, Lan\c{c}on et al.show that the relaxed configuration with $\delta z=8\;pm$ is energetically favorable. 
They conclude that $IDB^*$ as proposed by Northrup et al.\ is a meta-stable configuration and the actual $IDB^*$ ground state is only reached when an initial inter-domain shift is introduced. 
The findings of Lan\c{c}on et. al are experimentally corroborated by experimental measurements of inter-domain shifts in GaN nanowires containing IDBs, measured by Bragg Coherent X-ray imaging \cite{labat_inversion_2015,li_mapping_2020}.
Similarly, the relaxed $IDB^{\prime}$ structure, that we propose, could be sensitive picometer- scale inter-domain shifts to the initial configuration. Further studies are needed to assess this possibility.

\section{Conclusions}
\label{Sec:conclusions}
We have presented a notation to characterize IDBs based on the stacking sequence mismatch between the two polarity inverted domains. 
With the aid of this notation we were able to propose two IDB models that have not been previously reported on. 
We perform DFT calculations to estimate the domain wall energies of the IDB models and assess their energetic viability. 
Our calculations use defect supercells with periodic boundary conditions with typically two defect planes to recover the periodicity. 
However, for some planar defects two successive appearances of the defect planes fails to recover the periodicity. 
In such cases we devise an improved method to estimate domain wall energies by incorporating additional defect planes in the supercell. 
The IDB models we introduce and the accepted $SMB$ model require this method to estimate domain wall energy accurately. 
One of the IDB models we propose - $IDB^{\prime}$ - is energetically viable and has a domain wall energy that is lower than Holt-$IDB$ but higher than $IDB^*$.\\

We provide an alternate explanation for previous experimental interpretation of TEM imagery of $IDB^*$\textendash~$I_1\;SF$ interactions. 
We propose $IDB^{\prime}$ as the most suitable to describe the defect plane arising from this interaction over the previous interpretation that identified this defect as Holt-$IDB$. 
Additionally we analyze $IDB^*$ interacting with other SF models.
This analysis yields a possible scenario for the formation of Holt-$IDB$ via an $IDB^*$\textendash~$E\:SF$ interaction.\\

Further study is required to determine the electronic signature of the newly proposed $IDB^{\prime}$ model. 
Our calculations indicate that $IDB^{\prime}$ is not electronically inert, i.e. it produces states in the band-gap. 
However, the accuracy of our band-structure calculations are limited due to the use of PBE functionals and their well documented~\cite{perdew_density_1985} underestimation of band-gaps. 
An improved characterization of the changes in band-structure due to  $IDB^{\prime}$ can be achieved with DFT calculations employing hybrid functionals. 
Additionally, the $IDB^{\prime}$ and $IDB^{\prime\prime}$ models we proposed in this work are valid models for IDBs in other materials that have the wurzite crystal structure. 
Further studies are needed to assess the viability of these models in other wurzite materials.\\ 
\section*{Acknowledgments}
All computations for this work were performed on The Pennsylvania State University's Institute for CyberScience Advanced CyberInfrastructure (ICS-ACI) and the CyberLAMP computer clusters. 
CyberLAMP is funded by the National Science Foundation (NSF) under grant number 1626251. 
M.M.F.U. acknowledges training provided by the Computational Materials Education and Training (CoMET) NSF Research Traineeship (grant number DGE-1449785). 
The content of this paper is the sole responsibility of the authors and does not necessarily represent the views of ICS-ACI or the National Science Foundation.

\end{document}